# A Monte Carlo study of ligand-dependent integrin signal initiation

Federico Felizzi and Dagmar Iber[1]

D-BSSE, ETH Zurich, Mattenstrasse 26, CH-4058 Basel, Switzerland

[1]Author for correspondance, Dagmar.Iber@bsse.ethz.ch

**Running title:** Ligand-induced integrin signal initiation

**Abstract**

Integrins are allosteric cell adhesion receptors that control many important processes, including cell migration, proliferation, and apoptosis. Ligand binding activates integrins by stabilizing an integrin conformation with separated cytoplasmic tails, thus enabling the binding of proteins that mediate cytoplasmic signaling. Experiments demonstrate a high sensitivity of integrin signaling to ligand density and this has been accounted mainly to avidity effects. Based on experimental data we have developed a quantitative Monte Carlo model for integrin signal initiation. We show that within the physiological ligand density range avidity effects cannot explain the sensitivity of cellular signaling to small changes in ligand density. Src kinases are among the first proteins to be activated, possibly by trans auto-phosphorylation. We calculate the extent of integrin and ligand clustering as well as the speed and extent of Src kinase activation by trans auto-phosphorylation or direct binding at different experimentally monitored ligand densities. We find that the experimentally observed ligand density dependency can be reproduced if Src kinases are activated by trans auto-phosphorylation or some other mechanism limits integrin-dependent Src kinase activation. We propose that Src kinase and thus cell activation by trans auto-phosphorylation may provide a mechanism to enable ligand-density dependent responses at physiological ligand densities. The capacity to detect small differences in ligand density at a ligand density that is large enough to permit cell adhesion is likely to be important for haptotaxis.

## Introduction

Integrins are allosteric proteins that can respond to extracellular and intracellular stimuli and change their affinity for ligand (Hynes, 2002). The two extreme conformations, an 'open' and a 'closed' one, bind ligand with maximal and minimal affinity respectively. The extracellular conformational changes are accompanied by movements of the intracellular domains which lead to a separation of the integrin tails which enables integrins to bind to regulatory proteins inside the cell and to link to the cytoskeleton (Hynes, 2002). The 'open', active integrin conformation can be stabilized both by ligand binding to the extracellular part of integrins ('outside-in' signaling) and by binding of intracellular proteins to the cytoplasmic integrin tail ('inside-out' signaling).

Integrins have relatively short cytoplasmic tails without catalytic activity. Signal initiation therefore requires the activation of proteins that bind to the cytoplasmic domains of the separated tails (Banno and Ginsberg, 2008). Important early signaling proteins include Src kinases (Arias-Salgado et al., 2003; Su, 2008, p22862) and focal adhesion kinase (FAK). Active integrin-bound c-Src and FAK phosphorylate numerous targets which include important other signaling proteins that control cell spreading, cell migration, cell proliferation and apoptosis (Harrison, 2003).

Src kinases are particularly important for outside-in signaling but appear dispensable for inside-out signaling. Thus a synthetic peptide (RGT) that mimicks the c-Src binding motif and disrupts the interaction of $\alpha$IIb$\beta$3 with Src kinase selectively inhibits outside-in signaling but does not affect inside-out signaling in human platelets (Su et al., 2008). There are many different Src kinases that show different binding specificities. Thus while the related Src family kinases (SFKs) Yes, Hck, and Lyn bind to both $\beta$1 and $\beta$3 integrin tails c-Src and Fyn only bind to a C-terminal motif of the $\beta$3 tail (but not to the $\beta$1 tail) (Arias-Salgado et al., 2003 Obergfell, 2002, p36855).

Many details are known about the mechanism and the sequence of events by which integrins activate Src kinases. In the quiescent state the bulk of c-Src is clamped and inactive (Harrison, 2003). Phosphorylation of Tyr-529 (Tyr-527 in chicken (Takeya and Hanafusa, 1983)) by integrin-associated Csk-(C-terminal Src kinase) stabilizes the clamped conformation (Obergfell et al., 2002), and in the resting state 90-95% of all Src kinases are phosphorylated by Csk on Tyr527 (Roskoski, 2004). Src kinases can bind to the integrin tails already in the quiescent state and it has been suggested that this interaction renders Src kinases to be partially unclamped, or primed. Whether Src kinases bind also to the closed integrin conformation has not been resolved, but full Src activation requires the stabilization of the active integrin conformation (Obergfell et al., 2002).

Full c-Src activation and its "switching" to a stable activated state has been proposed to be achieved by trans autophosphorylation of the activation loop (Y418) which enables substrate access to the catalytic cleft (Sicheri and Kuriyan, 1997; Sicheri, 1997, p38371). Ligand-induced integrin clustering can increase the local c-Src concentration and thereby facilitate trans autophosphorylation. It has therefore come as a surprise that Src kinase was fully (if transiently) activated upon ligand binding also in the absence of talin (Zhang et al., 2008) even though integrins are not observed to cluster in the absence of the cytoplasmic protein talin (Saltel et al., 2009). Ligand binding may also contribute to the activation of c-Src by promoting dephosphorylation of Tyr-529 via tyrosine phosphatases (Su et al., 1999) and the dissociation of Csk from β3 (Obergfell et al., 2002). However, in the absence of talin there was also no detectable activation of β1 integrins 90 minutes after ligand exposure (Zhang et al., 2008).

Since the details and the relative contributions of the processes that affect Src kinase activation are difficult to resolve experimentally we developed a carefully parameterized, quantitative Monte Carlo simulation to explore the two discussed activation mechanisms, i.e. Src kinase activation upon binding to integrins (Fig. 1A) or by trans activation between neighbouring Src kinases (Fig. 1B). We expected that Src kinase activation by trans-activation would limit spurious, ligand-independent activation and might be inefficient unless ligand density is high and we introduce processes that enhance integrin clustering, including avidity effects or a positive feedback via talin.

To our surprise, for physiological parameter choices both mechanisms showed similar levels of spurious activation, and no externally enforced integrin clustering was necessary to enable Src kinase activation by trans-activation. Thus rapid Src kinase activation could be observed also without any supporting feedbacks (as may be provided by talin) and the transient activation of Src kinase in the absence of talin (Zhang et al., 2008) could be reproduced as long as a negative feedback was introduced as may be provided by the experimentally observed removal of Src binding sites by cleavage of the integrin tail (Flevaris et al., 2007).

Experiments reveal a strong ligand density dependency of integrin activation (Cox et al., 2001). Since the spacing between ligands is much more important than the average ligand density the remarkable sensitivity to ligand density has previously been accounted mainly to avidity effects and has been subject of many experimental and theoretical studies (i.e. (Koo et al., 2002 Ward, 1994, p981, Bunch, 2010, p35155, Carman, 2003, p11442)). While avidity effects enhance integrin-ligand binding and clustering in our simulation they are insufficient to explain the experimentally observed impact of ligand density on speed and extent of activation. This is because physiological / experimental ligand densities are high, and small reductions in ligand density therefore do not limit Src kinase activation unless most Src kinase interactions with integrins fail to trigger Src activation as is the case when Src kinases are activated by trans auto-phosphorylation. We therefore propose that Src kinase and thus cell activation by trans auto-phosphorylation provides a mechanism to enable ligand-density dependent responses at high ligand densities. High ligand densities are likely to be important for successful cell adhesion while the capacity to detect density differences is likely to be important for haptotaxis.

## Results

### Presentation of ligand

In experiments cell spreading is strongest when surfaces are covered with about 10 μg per ml fibronectin and is much slower but still detectable at 0.1 μg per ml fibronectin. According to measurements coverage of surfaces with 0.1, 1 or 10 μg per ml fibronectin results in surface densities of 1436, 3850, and 6500 epitope sites per $\mu m^2$ (Vitte et al., 2004). If we assume a dimension of 10x10 $nm^2$ per integrin-ligand binding site then there are 10000 potential integrin-ligand binding sites per $\mu m^2$. If the binding sites were as small as 5x5 $nm^2$ then there would be a maximum of 200x200=40000 integrin-ligand binding sites per $\mu m^2$. The grid in our simulation represents 1 $\mu m^2$ and accordingly has either 100x100 or 200x200 lattice sites. For 6500 epitope sites per $\mu m^2$ (grid) virtually all ligands are juxtaposed to some other ligand even if the binding site was as small as 5x5 $nm^2$ (Fig. 2 C,D). As the ligand density falls below 3850 per $\mu m^2$ (grid) the fraction of ligands that is juxtaposed falls dramatically, and in case of a larger grid an even smaller fraction of ligands is juxtaposed (Fig. 2 C,D, dashed lines).

### Ligand-dependent Integrin activation

The binding of ligands can stabilize the open integrin conformation and thereby enable integrin activation (Hynes, 2002). Since we focus on the initiation of the response and rapid Src kinase activation is observed also in talin knock-out cells (Zhang et al., 2008) we neglect talin and discuss the impact of a talin-dependent positive feedback on integrin activation elsewhere. If ligand is not limiting then in equlibrium the fraction of open integrins is given by

$$\frac{I_o}{I_{tot}} = \frac{K_C(1+L/K_D)}{1+K_C(1+L/K_D)}$$

Equation 1

Here L refers to the ligand concentration, $K_C$ reflects the conformational equilibrium in the absence of ligand, and $K_D$ refers to the integrin-ligand dissociation constant. In the absence of ligand only about 2-3% of all integrins are in the open, active conformation (Tadokoro et al., 2003), and we therefore set the conformational equilibrium constant to $K_C = 0.03$. The integrin-ligand affinities have been measured in solution, and in agreement with experiments we predict negligible integrin activation at concentrations that are typical for soluble ligand in body fluids and plasma (450-900 nM (Faull and Ginsberg, 1995) - marked by vertical bar in Fig. 2B) unless the open integrin conformation is stabilised by reagents such as antibodies or $Mn^{2+}$ (Fig. 2B, dashed black line). The conformational bias to the low affinity conformation thus protects integrins from being activated by soluble ligand in body fluids and plasma.

In experiments cell spreading is strongest when surfaces are covered with about 10 μg per ml fibronectin, which corresponds to about 6500 epitope sites per $\mu m^2$ (Vitte et al., 2004). If we assume that the binding distance between integrins on the cell surface and ligand on the substrate is about 20 nm then the maximal concentration of ligand in the reaction volume will be 0.5 mM. As discussed elsewhere (Iber and Campbell, 2006) experimental measurements suggest that the 2-dimensional affinity for the interaction of integrins with surface-bound ligand is much lower than the (volume-adjusted) 3-dimensional dissociation constant ($K_D = 0.3$ μM) (Bell, 1978; Kuo and Lauffenburger, 1993; Moy et al., 1999). Thus instead of full binding (Fig. 2B, grey, dashed vertical line) we predict partial ligand engagement of the $\alpha_V\beta_3$ integrins (Fig. 2B, shaded area). We note that such partial integrin activation is in good agreement with the experimental observations that integrin activation can still be enhanced with Mn2+ and antibodies that stabilize the open integrin conformation (i.e. (Cluzel et al., 2005)).

**Ligand-induced integrin clustering**

To calculate the expected number of juxtaposed ligand-bound integrins on a cell we carry out a Monte-Carlo simulation with ligands and integrins on a 100x100 grid which corresponds to about 1 $\mu m^2$ of cell membrane. Experiments report 1-30x$10^5$ integrins on fibroblasts and CHO cells (Akiyama and Yamada, 1985; Neff et al., 1982; Wiseman et al., 2004), and estimate the surface area of CHO cells as 1000-2500 $\mu m^2$ (Wiseman et al., 2004). The physiological integrin density may thus range between 40-3000 per $\mu m^2$, with a most likely range of 300-500 integrins per $\mu m^2$ (shaded area in Fig. 3B-D). Ligands are fixed while integrins can move by diffusion. Moreover, integrins can assume either a closed or an open conformation and bind ligand only in the open conformation (for details see Model section). If we use as binding energy $U_L$ = 1.35 kT to obtain 10% integrin-ligand binding (Fig. 3A), which marks the lower bound of the expected extent of integrin-ligand binding (Fig. 2B), then there are 4-5 juxtaposed ligand-bound integrins per $\mu m^2$ at high ligand densities (Fig. 3B). As the ligand density falls to 3850 epitope sites per $\mu m^2$ the number of juxtaposed ligand-bound integrins halves, and as the ligand density is reduced to 1436 ligands per $\mu m^2$ the number of juxtaposed ligand-bound integrins falls another 10-fold (Fig. 3C). We note that if the dimension of the integrin-ligand-binding site was smaller such that there were 40000 instead of 10000 binding sites per $\mu m^2$ then the figures would be some 80-fold lower. Avidity effects, i.e. a 4kT higher binding energy for the integrin-ligand interaction for juxtaposed integrins (Table 1), increase the number of juxtaposed ligand-bound integrins some 15-fold at high ligand densities. To achieve a similar number of juxtaposed ligand-bound integrins without avidity effects the affinity would need to be increased to $U_L$ = 3 kT which would then result in 40% rather than 10% ligand-bound integrins at 6500 ligands per $\mu m^2$ (Fig. 3D). A binding energy of $U_L$ = 3 kT thus marks a likely upper bound of integrin-ligand binding (Fig. 2B).

**Src kinase activation**

Two mechanisms for Src kinase activation have been proposed, i.e. upon binding to integrins (Fig. 1A) or by trans auto-phosphorylation (Fig. 1B). To explore these mechanisms for integrin-dependent Src kinase activation we introduce three activity states into our Monte Carlo simulation, a clamped state, $S_C$, an open state, $S_O$, and an active state, $S_A$. In the quiescent state 90-95% of Src kinases are clamped (Roskoski, 2004). Unless in the clamped state, Src kinases can interact with integrins and 3% of integrins are bound by Src kinases (Obergfell et al., 2002). It has not been resolved whether Src kinases can bind to both or only to the open integrin conformation (Obergfell et al., 2002) and we therefore consider both binding modes with Src kinases binding either restricted to the open integrin conformation (Fig 1, grey arrows) or not (Fig. 1, all arrows). Independent of whether Src kinases can bind only to the open integrin conformation or to all integrins and whether Src kinases are activated by trans autophosphorylation (Harrison, 2003) or directly by binding to integrins we observe full activation of Src kinases also in the absence of ligand (Fig. 4 A-D, dotted lines). Spurious Src kinase activation in the absence of ligand can be avoided if we either limit the likelihood of activation in the absence of ligand or introduce a large rate of Src kinase deactivation.

There is good experimental evidence for active processes that deactivate Src kinases. Thus experiments show that while Src kinases can be rapidly and fully activated within 5 minutes also in the absence of talin, activation is then transient and decays to about 50% of maximal activity within 15-20 minutes (Zhang et al., 2008). Other experiments show that the Src-dependent activation of Calpain leads to the proteolytic cleavage of the integrin tail, which removes the Src binding site (Flevaris et al., 2007). To reproduce these observations we introduce a Src deactivation step in our simulation such that removal of 70% of Src binding sites on integrin tails after 5000 MCS leads to the deactivation of one to two thirds of Src kinases as observed in experiments (Flevaris et al., 2007;

Zhang et al., 2008). We note that the same deactivation probability, $p_D=2x10^{-5}$ (Fig. 4F, black), could be used for all four possible Src kinase activation mechanisms (Fig. 4 E) without limiting the extent of Src kinase activation (Fig. 4G, black). While the removal of the Src kinase binding site provides a powerful mechanism to control Src kinase activity we note that a similar reduction in Src kinase activity could also have been achieved by increasing the probability of Src kinase deactivation some 15-fold.

Also with the deactivation process included there is, however, still considerable ligand-independent Src kinase activation (Fig. 4H). Ligand-independent Src kinase activation must therefore be reduced by lowering the probability of Src kinase activation, $p_A$, in the absence of ligand (Fig. 4H). If Src kinases are activated by trans auto-phosphorylation or by direct binding to open integrins the probability of activating Src kinases in the absence of ligand needs to be 0.01 or less ($p_A \leq 0.01$). Proteins that limit Src kinase activation (i.e. Csk) have indeed been described to be associated with integrins and to dissociate upon ligand binding (Obergfell et al., 2002). We note that if Src kinases could be activated by direct binding to any integrin then the probability of activating in the absence of ligand would need to be considerable smaller than the probability of deactivation ($p_A << p_D = 2x10^{-5}$) to achieve ligand-dependent Src kinase activation (Fig. 4H, black, dashed line). Since the probability of ligand-independent activation must be kept low also if Src kinases are activated by trans auto-phosphorylation we conclude in agreement with experimental studies (Obergfell et al., 2002) that Src kinase activation must be mainly the result of binding to the open integrin conformation. Binding of Src kinases to closed integrins sequesters Src kinases in unproductive interactions and accordingly the speed and extent of Src kinase activation are lower if Src kinases can bind also to closed integrin conformations (Fig. I-L). Notwithstanding these limitations on parameter choices, all four binding/activation mechanisms, however, enable a ligand-density dependent response (Fig. I-L).

**The impact of Src kinase and integrin densities and affinities on Src kinase activation**

The speed of Src kinase activation depends on the density of Src kinases and integrins as well as on the affinity of interaction. Activation is fastest for low Src densities, high integrin densities, and high affinities (Fig. 5A-D). For large Src densities or low integrin densities activation is slow because the number of juxtaposed integrins is limiting. Similarly, for low Src-integrin affinity there is insufficient binding to achieve rapid activation of Src kinases.

Within the physiological integrin density range Src kinase activation appears not very much affected by density changes (Fig 5 C,D, shaded area). A further reduction in integrin density leads, however, to a slower cellular response, and this may offer regulatory potential to the cell by integrin internalization. The physiological Src kinase density and the affinity of its interaction with integrins can be estimated from available experimental data. Thus in the quiescent state 90-95% of Src kinases are clamped (Roskoski, 2004). Unless in the clamped state, Src kinases can interact with integrins and 3% of integrins are bound by Src kinases (Obergfell et al., 2002). If the binding is restricted to the open integrin conformation then we find only a single parameter set that matches all experimental data, and we need to use an integrin-Src binding energy of 5 kT and a density of 300 Src kinases per $\mu m^2$ (Fig. 6 LHS, dotted lines). If Src kinases can bind to all integrin conformations then there is a range of Src kinase affinities (≤ 0.1 kT) and densities (≥ 300 Src kinases per $\mu m^2$) within which a lower affinity can be compensated for by a higher density and vice versa (Fig. 6 RHS). For best comparison of the two binding modes we use 300 Src kinases per $\mu m^2$ in both cases and an integrin-Src binding energy of 5 kT and 0.1 kT for binding to the open or any integrin conformations respectively. We note that such Src kinase density ensures rapid Src kinase activation independently of whether Src kinase binding is restricted to the open integrin conformation (Fig. 5 A,B).

**The impact of ligand density and avidity on Src kinase activation**

Both the speed and the extent of Src kinase activation are strongly affected by ligand density (Fig. 7A, B). Thus if we plot the average number of Monte Carlo steps (MCS) that is required to activate 50% of Src kinases in response to ligand binding then we observe faster responses at higher ligand density. At higher binding affinities the threshold is moved to lower ligand densities. We note that if Src kinases can bind to all integrin conformations then the speed and extent of Src kinase activation is lower (Fig. 7, RHS). This can be accounted to the low integrin-Src affinity (or Src kinase density) that we have to use to meet the experimentally observed low extent of Src kinase-integrin binding in the absence of ligand (Fig. 6).

For the likely physiological binding energies (that lead to 10-40% integrin binding at high ligand densities) and Src kinase activation by trans auto-phosphorylation the waiting time sharply increases and the extent of Src kinase activation decreases for ligand densities between 1436 and 3850 epitopes per $\mu m^2$ (Fig. 7 A, B, E, F). This correlates well with experimental results according to which cell adhesion is much slower and ineffecient if surfaces are coated with a 10-fold lower ligand concentration (which corresponds to a 2-fold lower density of coated epitopes). We note that this result follows without any further tuning of parameters, and all parameters have been calibrated carefully by comparison of simulation results with experimental data (Table 1).

If Src kinases are activated by trans auto-phosphorylation then avidity effects barely affect the ligand-density dependent speed (Fig. 7 A,B grey lines) and extent (Fig. 7 E,F grey lines) of Src kinase activation. On the other hand, if Src kinases are activated directly by integrin binding then avidity effects can slow down Src kinase activation (Fig 7C, D grey lines) and limit the extent of Src kinase activation (Fig. G,H grey lines) at lower ligand densities. However, the impact of avidity effects is insufficient to enable a discrimination of ligand density in the physiologically relevant density range unless we limit the probability of Src kinase activation upon binding to integrins to 10% when Src kinases can bind only to open integrins (Fig 7, C, G, dashed lines) and to 2% if Src kinases can bind to all integrins (Fig 7, D,H dashed lines).

We therefore conclude that discrimination of ligand densities at physiological ligand densities is possible only when Src kinase interactons with ligand-bound integrins do not always result in Src kinase activation.

**Discussion**

We have developed a quantitative, mechanistic Monte Carlo simulation to explore Src kinase activation during the initiation of intracellular signalling in response to ligand-dependent integrin activation. We carefully calibrated all parameters in the simulation by comparison of our simulation results to experimental data. Thus the conformational energies and the binding energies were set to reproduce the experimentally observed extent of ligand-independent and ligand-dependent integrin and Src kinase activation. The ligand and integrin densities have been determined in experiments. The only two parameters for which we did not have unambiguous experimental data were the integrin-Src kinase affinity and the Src kinase density. Depending on whether we permitted Src kinases to bind only to the open integrin conformation or to all integrin conformations we arrived at different choices for these parameters. However, our conclusions are not affected by these choices.

First of all we note that the physiological integrin and Src kinase densities appear to be balanced to enable fast and efficient activation of downstream signalling. Internalisation of integrins and calpain-dependent cleavage can therefore be expected to be powerful regulatory measures to control integrin-dependent Src kinase activation.

Integrin signalling is remarkably sensitive to ligand density in spite of the high local ligand density that is typically used in experiments and encountered *in vivo*. Experiments further show that cell adhesion is successful at lower average ligand densities if ligand is pre-clustered rather than homogenously distributed and this has been accounted to avidity effects (i.e. (Koo et al., 2002 Ward, 1994, p981, Bunch, 2010, p35155, Carman, 2003, p11442)). Our carefully parameterized simulation now reveals that avidity effects alone cannot explain the experimental data because avidity effects are not limited by small changes in ligand density in the physiological ligand density range. Such sensitivity can be gained only if less integrins are bound by ligand or integrin-dependent Src kinase activation is inefficient. However, the extent of integrin-ligand binding (10%) is low already in our simulation. Inefficient Src kinase activation could, in principle, be the consequence of regulatory proteins that also interact with integrins and limit Src kinase activation. It seems, however, that Src kinase activation by trans auto-phosphorylation can well serve such function.

Activation of Src kinases by trans auto-phosphorylation requires juxta-position of integrin-bound Src kinases for activation and thereby reduces the extent of Src kinase activation. If Src kinases are activated by trans-activation rather than directly upon integrin binding then the experimentally observed ligand density dependency can be reproduced without any parameter tuning or adjustments if Src kinases are activated by trans auto-phosphorylation. Src kinase activation by trans auto-phosphorylation thus provides an elegant mechanism to enable ligand density discrimination on the level of cellular signalling while permitting sufficient integrin-ligand interactions for cell adhesion. This may be important for haptotaxis as cells need to detect small differences in ligand density at a ligand density that is large enough to permit cell adhesion.

## Materials and Methods

### Monte Carlo Simulation

Simulations were initiated with the random placement of components on square grids ranging in size from 100x100 to 200x200 which were taken to represent 1 $\mu m^2$ of cell membrane. Unless otherwise stated we used 6500 ligands, 500 integrins and 300 Src kinases per grid. Ligands were fixed on the grid while unbound integrins and Src kinases could move to an unoccupied neighbour site in each simulation step. Binding was possible when players occupied the same lattice site.

Integrins can assume many different conformations (Hynes, 2002), but for simplicity we only considered the extreme conformations, a closed and an open one. In each simulation step integrins could change their conformation, i.e. switch from closed to open with probability $\exp(-U_C/kT)$ and switch back to the closed conformation with probability 1 unless bound by ligand. The conformational equilibrium was set such that in the absence of ligand only 2-3% of all integrins are in the open conformation (Tadokoro et al., 2003), i.e. $U_{Ic}$ = 3 kT. Since conformational changes are rapid compared to protein diffusion in a cell membrane we allow integrin to bind to ligand on the same lattice site regardless of integrin conformation. Integrins unbind ligand with probability $p=\exp(-U_L/kT)$. The binding energy $U_L$ = 1.35 kT corresponds to the 2-dimensional affinity (10% ligand binding at 6500 ligands per $\mu m^2$) of $\alpha_V\beta_3$ integrins ($1/K_D$ = 0.3 $\mu M$ , Plow and Ginsberg, 1981). We also considered a higher affinity $U_L$ = 3 kT which corresponds to 40% binding, and an avidity effect that enhances the affinity of binding for neighbouring integrins by $U_A$ = 4kT, with base affinity $U_L$ = 0.4 kT ($U_{L+A}$ = 4.4 kT, 10% binding) and $U_L$ = 1.35 kT ($U_{L+A}$ = 5.35 kT, 40% binding). Upon unbinding, integrins assume either the closed or the open conformation, with the probability of each conformation being determined by the equilibrium conformational ratio in the absence of ligand.

For the interaction between Src kinases and integrins we explore two binding scenarios. Src kinases can either bind only to open integrins or to all integrin conformations. In both cases ligand-bound integrins are bound with probability p=1 while unbound integrins are bound with probability p=0.01. Src kinases that can bind only to open integrins unbind with probability $p=\exp(-U_S/kT)$, $U_S = 5$ kT while Src kinases that can bind to any integrin conformation unbind with probability $p=\exp(-U_S/kT)$, $U_S = 0.1$ kT. In the resting state 90-95% of all Src kinases are inactive (Roskoski, 2004), and we therefore have for the conformational equilibrium energy between the clamped and the open Src kinase conformation $U_{Sc} = 3$ kT. The tyrosine kinase c-Src phosphorylates its optimal peptide substrates with $K_m = (k_{off} + k_{cat})/ k_{on} = 33$ μM (Songyang et al., 1995) and accordingly phosphorylation is immediate when two Src-bound integrins are juxtaposed.

**Acknowledgements**

This research was supported by a SystemsX grant as part of the RTD InfectX.

## References

Akiyama, S.K., and K.M. Yamada. 1985. The interaction of plasma fibronectin with fibroblastic cells in suspension. *The Journal of biological chemistry*. 260:4492-500.

Arias-Salgado, E.G., S. Lizano, S. Sarkar, J.S. Brugge, M.H. Ginsberg, and S.J. Shattil. 2003. Src kinase activation by direct interaction with the integrin beta cytoplasmic domain. *Proceedings of the National Academy of Sciences of the United States of America*. 100:13298-302.

Arias-Salgado, E.G., S. Lizano, S.J. Shattil, and M.H. Ginsberg. 2005. Specification of the direction of adhesive signaling by the integrin beta cytoplasmic domain. *J Biol Chem*. 280:29699-707.

Banno, A., and M.H. Ginsberg. 2008. Integrin activation. *Biochem Soc Trans*. 36:229-34.

Barsukov, I.L., A. Prescot, N. Bate, B. Patel, D.N. Floyd, N. Bhanji, C.R. Bagshaw, K. Letinic, G. Di Paolo, P. De Camilli, G.C.K. Roberts, and D.R. Critchley. 2003. Phosphatidylinositol phosphate kinase type 1gamma and beta1-integrin cytoplasmic domain bind to the same region in the talin FERM domain. *The Journal of biological chemistry*. 278:31202-9.

Bell, G.I. 1978. Models for the specific adhesion of cells to cells. *Science*. 200:618-627.

Calderwood, D.A., B. Yan, J.M. de Pereda, B.G. Alvarez, Y. Fujioka, R.C. Liddington, and M.H. Ginsberg. 2002. The phosphotyrosine binding-like domain of talin activates integrins. *J Biol Chem*. 277:21749-58.

Cluzel, C., F. Saltel, J. Lussi, F. Paulhe, B.A. Imhof, and B. Wehrle-Haller. 2005. The mechanisms and dynamics of (alpha)v(beta)3 integrin clustering in living cells. *J Cell Biol*. 171:383-92.

Cox, E.A., S.K. Sastry, and A. Huttenlocher. 2001. Integrin-mediated adhesion regulates cell polarity and membrane protrusion through the Rho family of GTPases. *Molecular Biology of the Cell*. 12:265-77.

Di Paolo, G., L. Pellegrini, K. Letinic, G. Cestra, R. Zoncu, S. Voronov, S. Chang, J. Guo, M.R. Wenk, and P. De Camilli. 2002. Recruitment and regulation of phosphatidylinositol phosphate kinase type 1 gamma by the FERM domain of talin. *Nature*. 420:85-9.

Faull, R.J., and M.H. Ginsberg. 1995. Dynamic regulation of integrins. *Stem Cells*. 13:38-46.

Flevaris, P., A. Stojanovic, H. Gong, A. Chishti, E. Welch, and X. Du. 2007. A molecular switch that controls cell spreading and retraction. *J Cell Biol*. 179:553-65.

Goksoy, E., Y.-Q. Ma, X. Wang, X. Kong, D. Perera, E.F. Plow, and J. Qin. 2008. Structural basis for the autoinhibition of talin in regulating integrin activation. *Molecular Cell*. 31:124-33.

Gonzalez, A.M., R. Bhattacharya, G.W. deHart, and J.C.R. Jones. 2010. Transdominant regulation of integrin function: mechanisms of crosstalk. *Cell Signal*. 22:578-83.

Harrison, S.C. 2003. Variation on an Src-like theme. *Cell*. 112:737-40.

Hilgemann, D.W. 2007. Local PIP(2) signals: when, where, and how? *Pflugers Arch*. 455:55-67.

Hynes, R.O. 2002. Integrins: bidirectional, allosteric signaling machines. *Cell*. 110:673-87.

Iber, D., and I.D. Campbell. 2006. Integrin activation - the importance of a positive feedback. *Bull Math Biol*. 68:945-956.

Koo, L.Y., D.J. Irvine, A.M. Mayes, D.A. Lauffenburger, and L.G. Griffith. 2002. Co-regulation of cell adhesion by nanoscale RGD organization and mechanical stimulus. *J Cell Sci*. 115:1423-33.

Kuo, S., and D. Lauffenburger. 1993. Relationship between receptor/ligand binding affinity and adhesion strength. *Biophys J*. 65:2191-2200.

Ling, K., R.L. Doughman, A.J. Firestone, M.W. Bunce, and R.A. Anderson. 2002. Type I gamma phosphatidylinositol phosphate kinase targets and regulates focal adhesions. *Nature*. 420:89-93.

Moy, V., Y. Jiao, T. Hillmann, H. Lehmann, and T. Sano. 1999. Adhesion energy of receptor-mediated interaction measured by elastic deformation. *Biophys J*. 76:1632-1638.

Neff, N.T., C. Lowrey, C. Decker, A. Tovar, C. Damsky, C. Buck, and A.F. Horwitz. 1982. A monoclonal antibody detaches embryonic skeletal muscle from extracellular matrices. *J Cell Biol*. 95:654-66.

Obergfell, A., K. Eto, A. Mocsai, C. Buensuceso, S.L. Moores, J.S. Brugge, C.A. Lowell, and S.J. Shattil. 2002.

Coordinate interactions of Csk, Src, and Syk kinases with [alpha]IIb[beta]3 initiate integrin signaling to the cytoskeleton. *J Cell Biol*. 157:265-75.

Pampori, N., T. Hato, D.G. Stupack, S. Aidoudi, D.A. Cheresh, G.R. Nemerow, and S.J. Shattil. 1999. Mechanisms and consequences of affinity modulation of integrin alpha(V)beta(3) detected with a novel patch-engineered monovalent ligand. *The Journal of biological chemistry*. 274:21609-16.

Plow, E.F., and M.H. Ginsberg. 1981. Specific and saturable binding of plasma fibronectin to thrombin-stimulated human platelets. *The Journal of biological chemistry*. 256:9477-82.

Roskoski, R. 2004. Src protein-tyrosine kinase structure and regulation. *Biochem Biophys Res Commun*. 324:1155-64.

Saltel, F., E. Mortier, V.P. Hytönen, M.-C. Jacquier, P. Zimmermann, V. Vogel, W. Liu, and B. Wehrle-Haller. 2009. New PI(4,5)P2- and membrane proximal integrin-binding motifs in the talin head control beta3-integrin clustering. *J Cell Biol*. 187:715-31.

Shattil, S.J. 2005. Integrins and Src: dynamic duo of adhesion signaling. *Trends in Cell Biology*. 15:399-403.

Sicheri, F., and J. Kuriyan. 1997. Structures of Src-family tyrosine kinases. *Curr Opin Struct Biol*. 7:777-85.

Songyang, Z., K.L. Carraway, M.J. Eck, S.C. Harrison, R.A. Feldman, M. Mohammadi, J. Schlessinger, S.R. Hubbard, D.P. Smith, and C. Eng. 1995. Catalytic specificity of protein-tyrosine kinases is critical for selective signalling. *Nature*. 373:536-9.

Su, J., M. Muranjan, and J. Sap. 1999. Receptor protein tyrosine phosphatase alpha activates Src-family kinases and controls integrin-mediated responses in fibroblasts. *Current biology : CB*. 9:505-11.

Su, X., J. Mi, J. Yan, P. Flevaris, Y. Lu, H. Liu, Z. Ruan, X. Wang, N. Kieffer, S. Chen, X. Du, and X. Xi. 2008. RGT, a synthetic peptide corresponding to the integrin beta 3 cytoplasmic C-terminal sequence, selectively inhibits outside-in signaling in human platelets by disrupting the interaction of integrin alpha IIb beta 3 with Src kinase. *Blood*. 112:592-602.

Suehiro, K., J. Mizuguchi, K. Nishiyama, S. Iwanaga, D.H. Farrell, and S. Ohtaki. 2000. Fibrinogen binds to integrin alpha(5)beta(1) via the carboxyl-terminal RGD site of the Aalpha-chain. *J Biochem*.

128:705-10.

Tadokoro, S., S.J. Shattil, K. Eto, V. Tai, R.C. Liddington, J.M. de Pereda, M.H. Ginsberg, and D.A. Calderwood. 2003. Talin binding to integrin beta tails: a final common step in integrin activation. *Science*. 302:103-6.

Takeya, T., and H. Hanafusa. 1983. Structure and sequence of the cellular gene homologous to the RSV src gene and the mechanism for generating the transforming virus. *Cell*. 32:881-90.

Vitte, J., A.-M. Benoliel, P. Eymeric, P. Bongrand, and A. Pierres. 2004. Beta-1 integrin-mediated adhesion may be initiated by multiple incomplete bonds, thus accounting for the functional importance of receptor clustering. *Biophysical Journal*. 86:4059-74.

Wiseman, P.W., C.M. Brown, D.J. Webb, B. Hebert, N.L. Johnson, J.A. Squier, M.H. Ellisman, and A.F. Horwitz. 2004. Spatial mapping of integrin interactions and dynamics during cell migration by image correlation microscopy. *J Cell Sci*. 117:5521-34.

Xi, X., P. Flevaris, A. Stojanovic, A. Chishti, D.R. Phillips, S.C.T. Lam, and X. Du. 2006. Tyrosine phosphorylation of the integrin beta 3 subunit regulates beta 3 cleavage by calpain. *J Biol Chem*. 281:29426-30.

Zhang, X., G. Jiang, Y. Cai, S.J. Monkley, D.R. Critchley, and M.P. Sheetz. 2008. Talin depletion reveals independence of initial cell spreading from integrin activation and traction. *Nat Cell Biol*. 10:1062-8.

**Figure Legends**

**Figure 1. A model for the ligand-dependent initiation of integrin signaling.**

A scheme of the simulated interactions if (A) Src kinases are activated upon binding to integrins or (B) by transactivation. Black arrows apply when Src kinases can bind to all and not just to the open integrin conformation. For details see text.

**Figure 2. The impact of ligand density on ligand clustering and integrin activation.**

**(A)** A typical grid with clustered (black) and non-clustered (grey) ligands. **(B)** The fraction of active $\alpha_V\beta_3$ integrins dependent on ligand concentration (L) and affinity ($K_d$) according to Eq 1 with conformational equilibriums constant $K_c$ = 0.03 (solid line) or when the open conformation is stabilised ($K_c$ = 1 - dashed line). The grey lines mark binding levels when ligand is either encountered in soluble form in the plasma and other body fluids (vertical bar), or ligand is presented on surfaces and either the lower 2D affinity (shaded area) or the 3D affinity (dashed line) is employed. **(C)** The expected number and **(D)** the fraction of ligands that is juxtaposed on a 100x100 (solid line) and a 200x200 grid (dashed line) for ligand densities as indicated.

**Figure 3. Ligand-induced integrin clustering.**

**(A)** The expected fraction of ligand-bound integrins with 4 different integrin-ligand binding modes: The binding energy is either (1) $U_L$ = 1.35 kT (solid black line); (2) $U_L$ = 0.4 kT but enhanced when integrins are juxtaposed ($U_{L+A}$ = 4.4 kT, solid grey line), (3) $U_L$ = 3.00 kT (dashed black line), and (4) $U_L$ = 1.35 kT, but enhanced when integrins are juxtaposed ($U_{L+A}$ = 5.35 kT, dashed grey line). **(B)** The expected number of juxtaposed integrins per cell with ligand-integrin affinity $U_L$ = 1.35 kT (1). The cellular integrin density is 100 (dotted line), 300, 500 (solid lines), or 1000 per μm2 (dashed line) and the ligand density is as indicated. The physiological integrin density range is shaded. **(C)** The expected number of juxtaposed integrin, with a cellular integrin density of 500 per $\mu m^2$ and the 4 different binding modes as described in (A).

**Figure 4. Src kinase activation and deactivation. (A-D)** Fraction of active Src kinases over time (Monte-Carlo steps) when Src kinase activation is not limited ($p_A = 1$) and there is no deactivation ($p_D = 0$). Src kinases are activated (A,B) upon integrin binding or (C,D) by trans auto-phosphorylation and Src kinases can bind to (A,C) open integrins or (B,D) to all integrins. Lines represent simulations with 6500 ligands per $\mu m^2$ (black solid line), 3850 (dotted grey line), 1438 (solid grey line) or in the absence of ligand (dotted black line). **(E)** Fraction of active Src kinases over time (Monte-Carlo steps) when Src kinases are deactivated and the Src binding site was removed from 70% of integrin tails after 5000 MCS. Src kinases are activated upon integrin binding (grey line) or by trans auto-phosphorylation (black line) and Src kinases can bind to open integrins (solid line) or to all integrins (broken line). **(F-G)** Parameter screening for the probability of Src kinase activation in the absence of ligand and the probability of Src kinase deactivation. The white asterix indicates the parameter set used in the simulation. **(F)** The fraction of active Src kinases in the presence of ligand. Shaded areas indicate parameter sets for which the fraction of active Src kinases exceeds 80% (black) or 50% (grey). **(G)** Deactivation of Src kinases by removal of 70% of all binding sites. The black area indicates the parameter set for which the fraction of active Src kinases returns to one to two thirds **(H)** The fraction of active Src kinases in the absence of ligand versus the activation probability, $p_A$. Src kinases are activated upon integrin binding (grey line) or by trans auto-phosphorylation (black line) and Src kinases can bind to open integrins (solid line) or to all integrins (broken line). **(I-L)** Fraction of active Src kinases over time (Monte-Carlo steps) when Src kinase activation is limited ($p_A \neq 1$) and there is deactivation ($p_D = 2x10^{-5}$). Src kinases are activated (I,J) upon integrin binding or (K,L) by trans auto-phosphorylation and Src kinases can bind to (I,K) open integrins or (J,L) to all integrins. Lines represent simulations with 6500 ligands per $\mu m^2$ (black solid line), 3850 (dotted grey line), 1438 (solid grey line) or in the absence of ligand (dotted black line).

**Figure 5. Density dependency of Src kinase activation.** Number of Monte Carlo steps (MCS) to achieve 50% Src kinase activation dependent on **(A,B)** Src density and **(C,D)** integrin density. Src kinases could bind either only to open integrins (LHS) or to all integrins (RHS). The lines represent the 4 binding modes as described in Figure 3: (1) $U_L$ = 1.35 kT (solid black); (2) $U_L$ = 3 kT (dashed black), or 4kT affinity enhancement for juxtaposed integrins and base binding energy (3) $U_L$ = 0.4 kT (solid grey) and (4) $U_L$ = 1.35 kT (dashed grey).

**Figure 6. Src kinase regulation in the absence of ligand.**

**(A-B)** Fraction of Src kinases in a clamped, inactive state and **(C-D)** fraction of Src-bound integrins dependent on the Src density for different integrin-Src binding affinities. (LHS )Src kinases could bind only to open integrins, and binding energies were set to $U_S$ = 5 kT (dotted line), $U_S$ = 6 kT (dashed line) and $U_S$ =7 kT (solid line). (RHS) Src kinases could bind to all integrins with binding energies $U_S$ = 0.05 kT (dotted line), $U_S$ =0.1 kT (dashed line) and $U_S$ =0.25 kT (solid line).

**Figure 7. The impact of ligand density and binding affinity of the speed and extent of Src kinase activation. (A-D)** Time (Monte Carlo steps) to 50% Src kinase activation and **(E-H)** steady state fraction of active Src kinases dependent on ligand density. Src kinases could bind either only to open integrins (LHS) or to all integrin conformations (RHS). (**A, B, E, F)** Src kinase activation by trans-activation (Fig. 1B). The lines represent the 4 binding modes as described in Figure 3 and 5: (1) $U_L$ = 1.35 kT (solid black); (2) $U_L$ = 3 kT (dashed black), or 4kT affinity enhancement for juxtaposed

integrins and base binding energy (3) $U_L = 0.4$ kT (solid grey) and (4) $U_L = 1.35$ kT (dashed grey). **(C, D, G, H)** Src kinase activation upon binding to integrins (Fig. 1A) with 10% integrin-ligand binding at high ligand density, i.e. $U_L = 1.35$ kT (black lines) or $U_L = 0.4$ kT with 4kT affinity enhancement for juxtaposed integrins (grey lines). Src kinase activation was either limited, $p_A \ll 1$ (dashed lines), or not, $p_A=1$ (solid lines).

**Tables**

**Table 1. Parameters employed in the simulation.**

| Symbol | Value | Reaction | Comment | References |
|---|---|---|---|---|
| $U_C$ | 3 kT | $I_c \leftrightarrow I_o$ | 2-3% open integrins in the absence of ligand | Tadokoro et al., 2003 |
| $U_{L+A}$ | 1.35 - 5.35 kT | $I_o \leftrightarrow I_oL$ | 10-40% integrin-ligand binding with and without avidity effects | Bell, 1978; Kuo and Lauffenburger, 1993; Moy et al., 1999 |
| $U_S$ | 5 kT / 0.1 kT | $I \leftrightarrow IS$ | Src kinase binding to open / all integrins | Obergfell et al., 2002 |
| $U_{SC}$ | 3 kT | $S_c \leftrightarrow S_o$ | 90-95% of Src kinases are clamped in the absence of ligand | Roskoski, 2004 |
| $p_A$ | ≤ 0.01 | $IS_o \rightarrow IS_a$ | probability of Src kinase activation in the ab-sence of ligand (per MCS) | Obergfell et al., 2002 |
| $p_D$ | $2 \times 10^{-5}$ | $S_a \rightarrow S_o$ | probability of Src kinase deactivation (per MCS) | Zhang et al., 2008; Flevaris et al., 2007 |

**Figures**

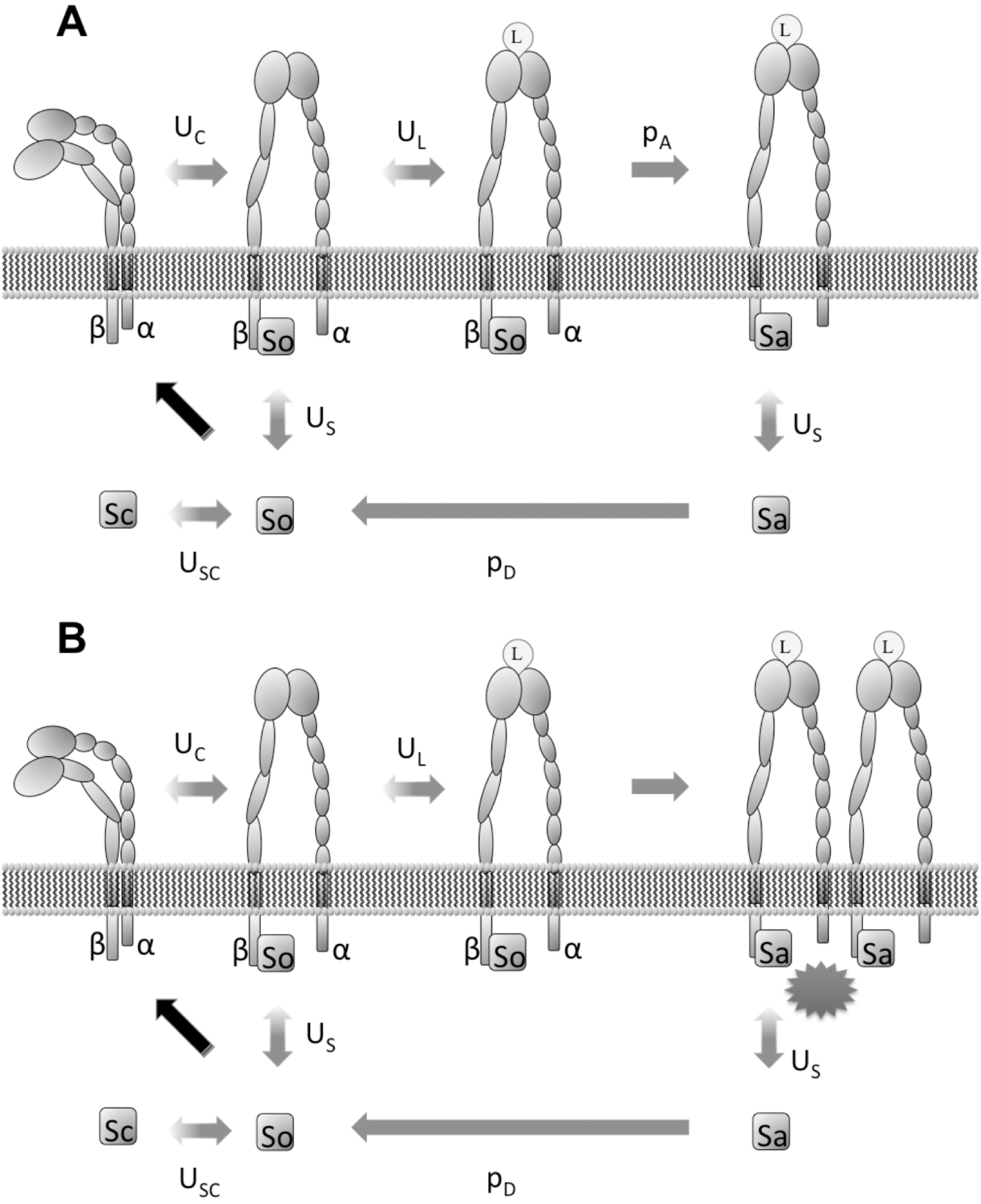

**Figure 1**

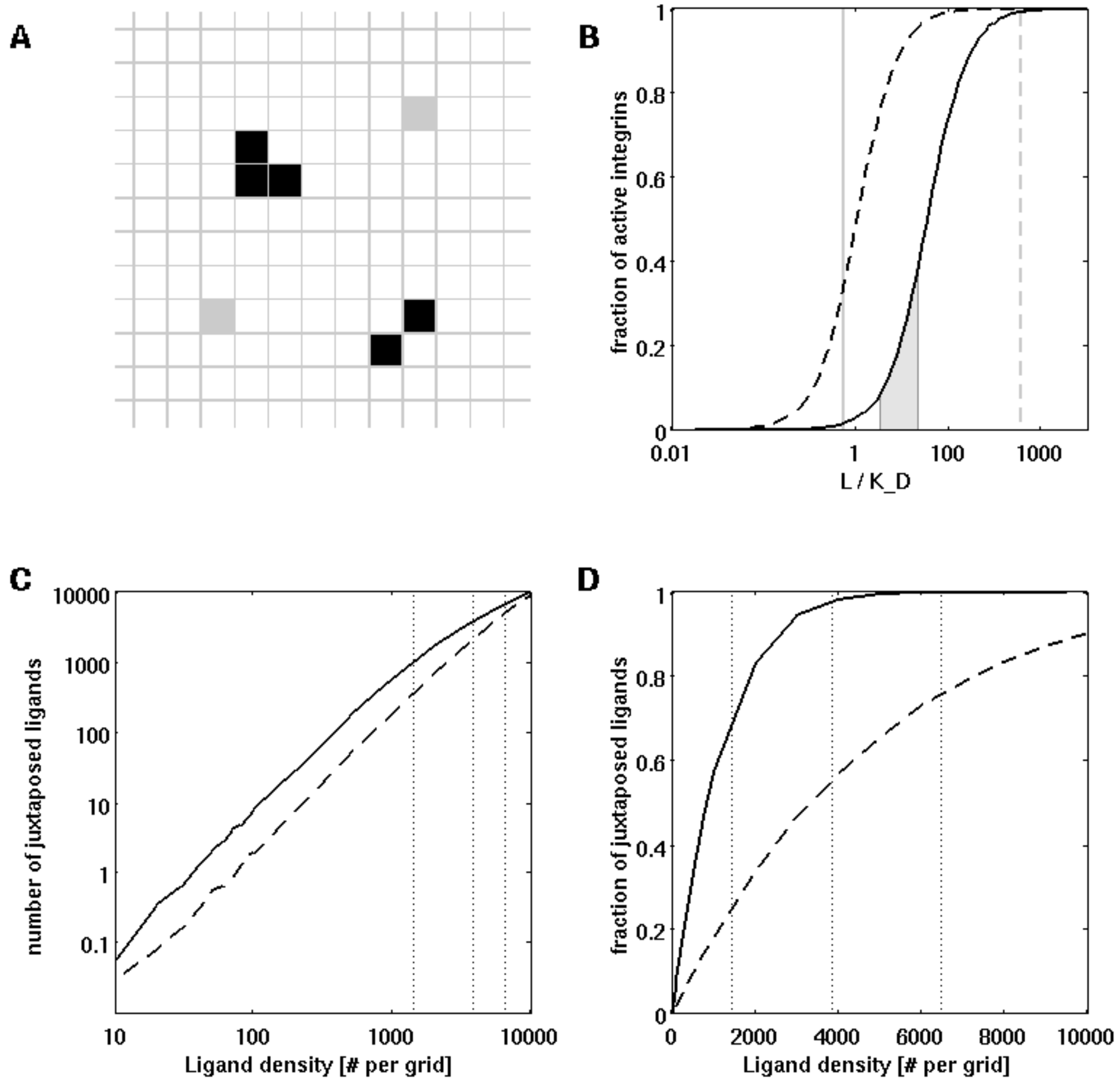
A
B
fraction of active integrins
L / K_D
C
number of juxtaposed ligands
Ligand density [# per grid]
D
fraction of juxtaposed ligands
Ligand density [# per grid]

**Figure 2**

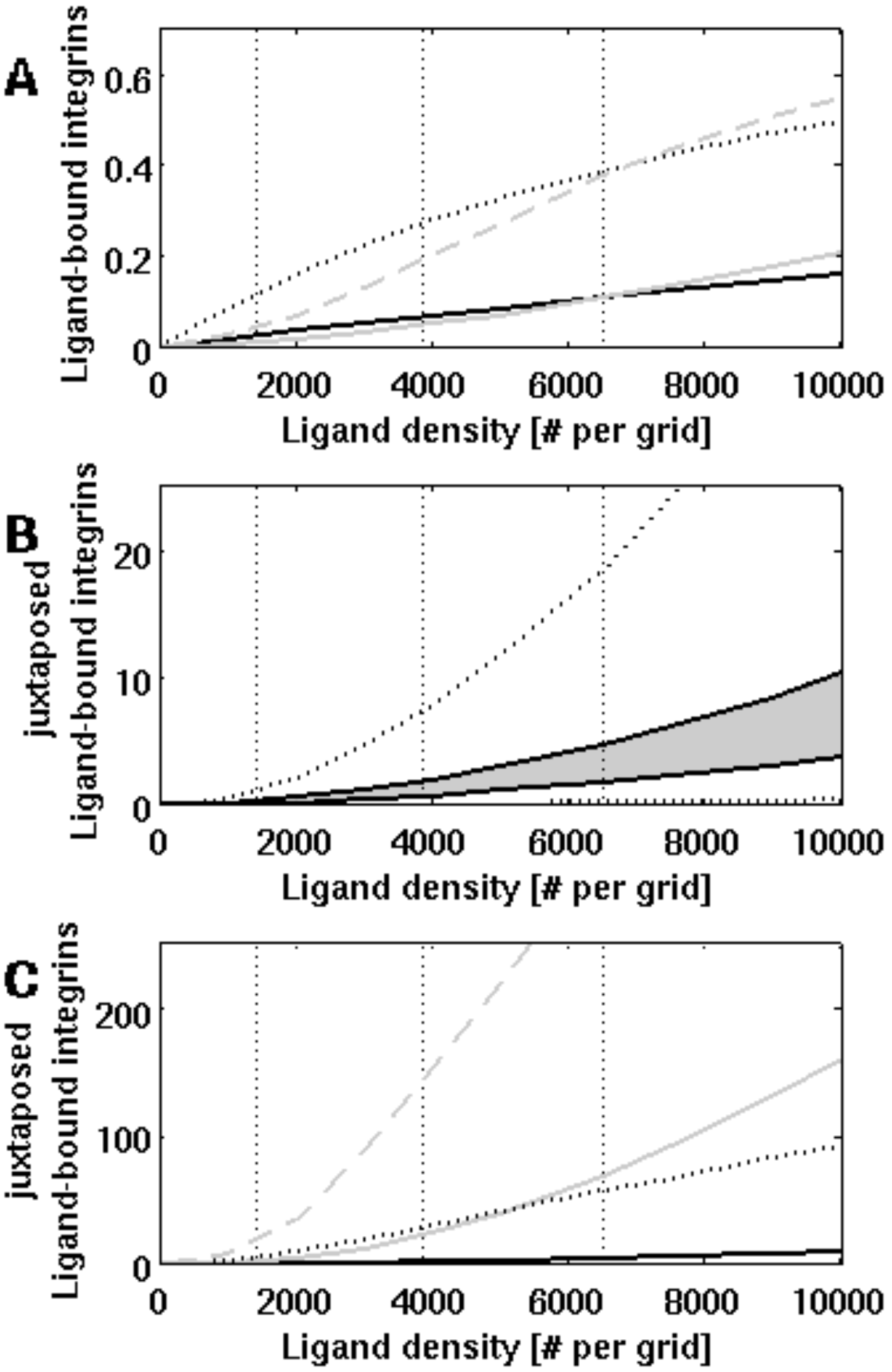

**Figure 3**

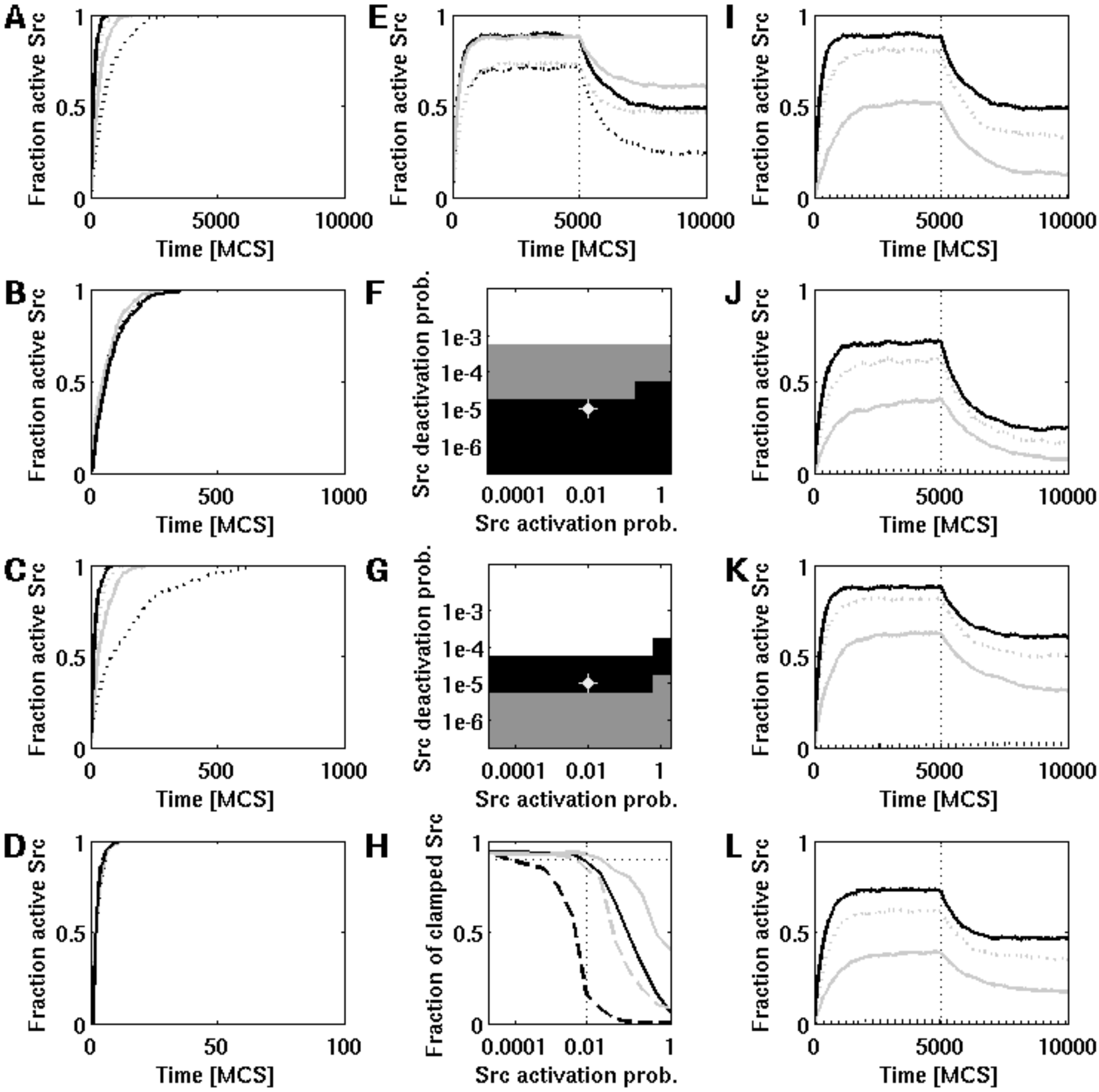

**Figure 4**

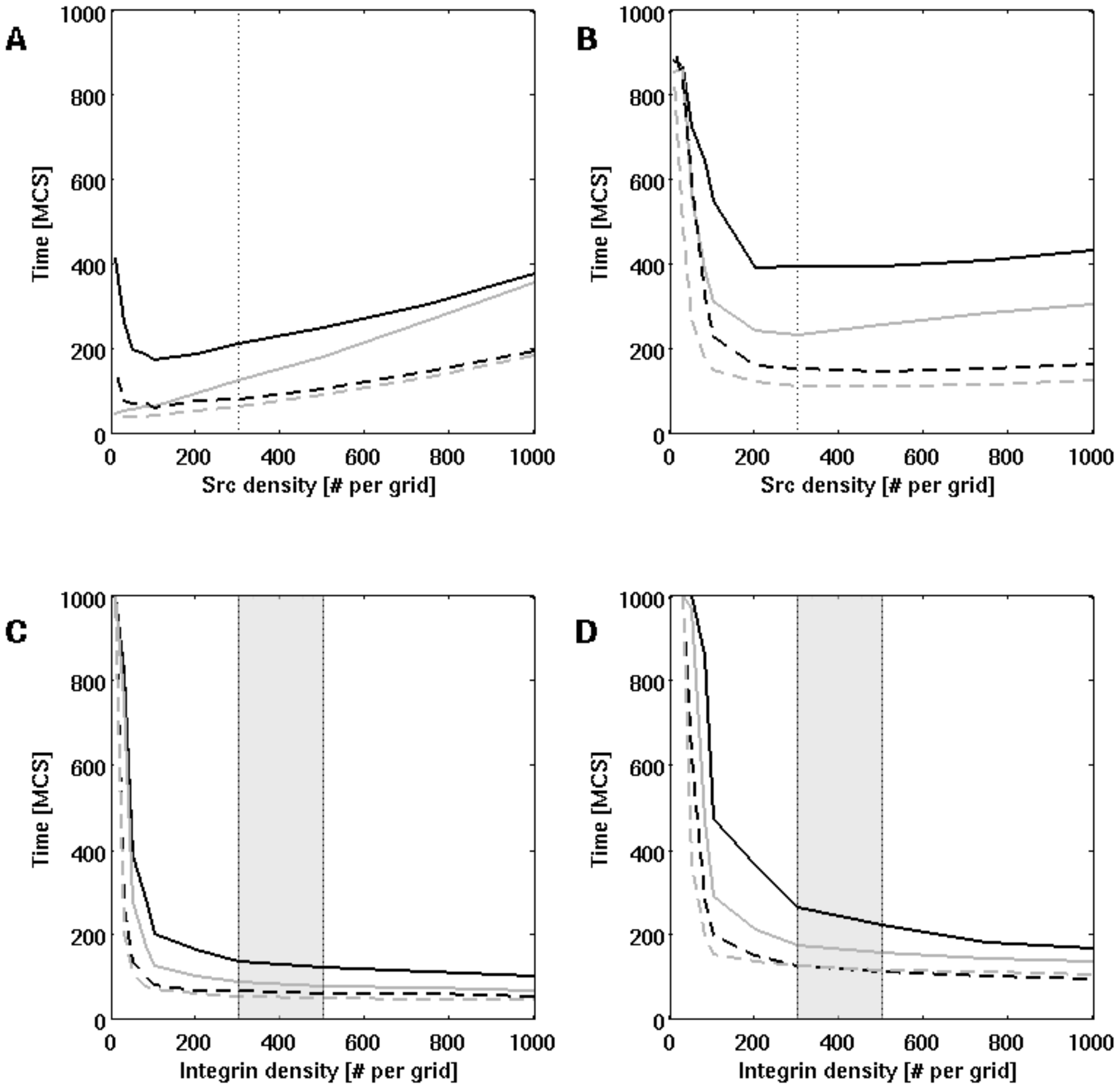

Figure 5

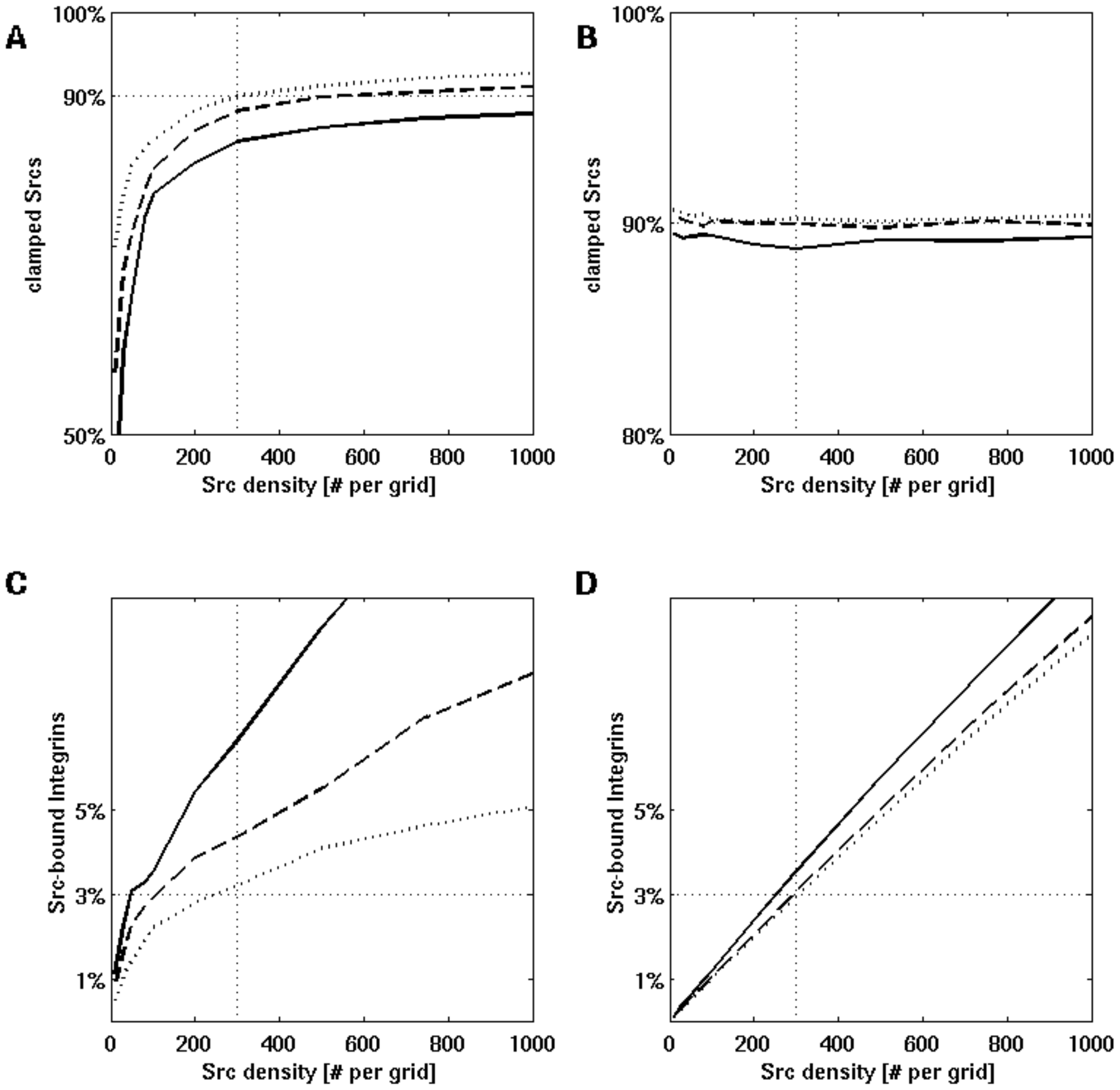

**Figure 6**

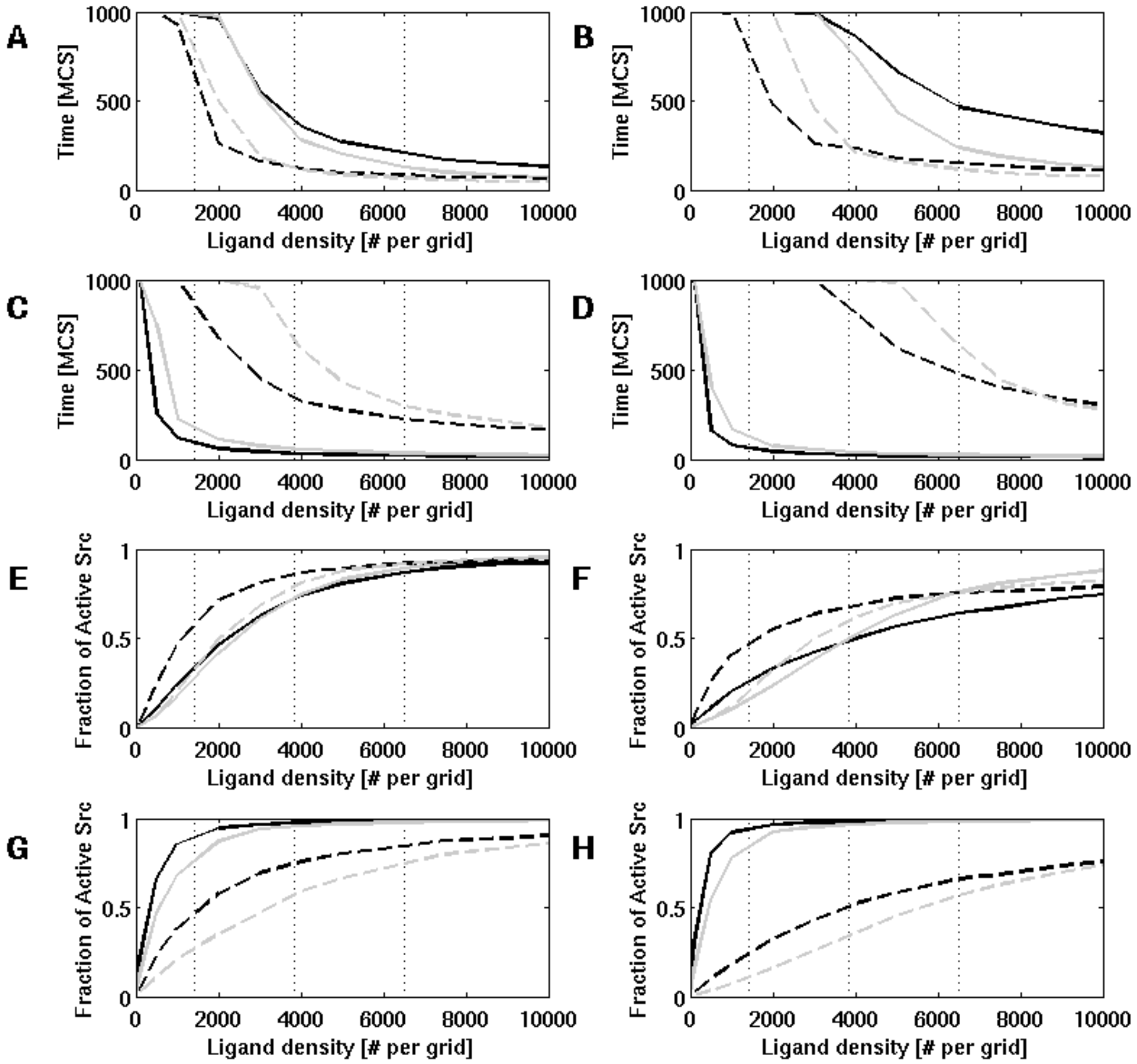

**Figure 7**